# Covalent functionalization of strained graphene

## Danil W. Boukhvalov*[a] and Young-Woo Son[a]

*Enhancement of the chemical activity of graphene is evidenced by first-principles modelling of chemisorption of the hydrogen, fluorine, oxygen and hydroxyl groups on strained graphene. For the case of negative strain or compression, chemisorption of the single hydrogen, fluorine or hydroxyl group is energetically more favourable than those of their pairs on different sublattices. This behaviour stabilizes the magnetism caused by the chemisorption being against its destruction by the pair formations. Initially flat, compressed graphene is shown to buckle spontaneously right after chemisorption of single adatoms. Unlike hydrogenation or fluorination, the oxidation process turns from the endothermic to exothermic for all types of the strain and depends on the direction of applied strains. Such properties will be useful in designing graphene devices utilizing functionalization as well as mechanical strains.*

## Introduction

Graphene is the novel and prospective material for future electronic[1–4] and spintronic devices,[5,6] solar cells, optical applications,[7,'8] supercapacitors and composite materials.[9-12] For those applications, chemical functionalization is routinely used and could change the electronic,[13-26] magnetic,[22-27] mechanical[20, 21,28,29] and chemical[30-32] properties of graphene significantly. For example, uniform one-sided functionalization of graphene[33] is nesessary for realizing of semiconducting behavior with high carrier mobility. In addition to chemical functionalization, changes of crystal structure of graphene such as corrugations[34,35] or formation of nanoribbons drastically change electronic structure[36–39] of graphene and provides enhancement of its chemical activity.[13,16,27]

Another kind of distortion of the graphene lattice is strain. Recent theoretical work predict unusual electronic properties of strained graphene.[40-44] Graphene usually have non-uniform contacts to the surface so that the routine heat treatment for device preparation produce of local strain on graphene. The role of strain in the migration of hydrogen atoms,[45] chemical activity of lithium adatoms,[46] electronic structure of graphene totally or half covered by hydrogen[47] and physical properties of corrugated graphene with partial hydrogenation[49] have been discussed in recent works. Previous works of one of the authors[13,50] discussed several energetically favourable configurations such as uniform coverage of one or both sides. However, such cases may not be fabricated at realistic conditions due to energy costs of early steps of functionalization. Thus, the systematic modelling of the possible adsorption behaviours of chemical species commonly used for graphene functionalization under the strain is necessary.

To understand the main principles of strained graphene functionalization, herein, the atomistic modelling on absorption of a single hydrogen (see Figure 1a), fluorine, oxygen adatom, hydroxyl group and its pairs has been performed. Hydrogen was chosen as the most detailed theoretically (see ref. 13 and references therein) and experimentally[17–19] studied chemical species for graphene functionalization. Other substitution of the hydrogen for the functionalization is fluorine. The higher stability of the totally and partially fluorinated graphene was reported in recent experimental[20,21] and theoretical[51,52] work. Graphene oxide could be described as graphene functionalized by oxygen (epoxy groups, Figure 1f-i) and hydroxyl groups (see ref. 14 and references therein). For the case of a pair adsorption, we have only considered that the two species are close to each other because the previous works have shown that, for pure,[13,'14] defected[27] and distorted[16] graphene, such a pairing is more energetically favourable than other cases. The changes of their cohesive energy due to the lattice deformation are interesting for the probable usage of strain effect for graphene oxide total reduction. The next step of our survey of the strained graphene functionalization is studying of the most common pairs of adsorbed species to check the probable scenarios of further functionalization: clusters,[18,19,53] lines or uniform formation of chemisorbed molecules or atoms.

## Results and Discussion

### Hydrogen, fluorine and hydroxyl groups chemisorption on strained graphene

Expansion of graphene provides decreasing of the chemisorption energy for all configurations of hydrogen adatoms (Figure 2), that is, the chemical activity of graphene is enhanced. Hereafter, the positive strain corresponds to the expansion of graphene while the negative to the compression. The cause of such enhancement with positive strains is an increase of the total energy of the graphene supercell defined above. This behavior is similar to the cases of graphene nanoribbons[26] and corrugated graphene.[16] For all types of positive strain (uniaxial strains perpendicular to the armchair/zigzag direction and isotropic strain) chemisorption energy remains positive: the chemisorption of the pair of hydrogen atoms in the *ortho* position from both sides of graphene sheet (Figure 1c) is always the most energetically favorable, while chemisorption of the single hydrogen is always energetically less favorable. Since the increase rate of the total energy of the graphene supercell with increasing positive isotropic strain is bigger than one with uniaxial strain (Figure 1j), the chemisorptions energy for this case decreases much faster as the magnitude of strain increases compared to those under uniaxial ones (Figure 2c). Here, we note that the preferred paired chemisorptions indicate an unstable local magnetism.

[a] Prof. dr. D. W. Boukhvalov, Prof. dr. Young-Woo Son
School of Computational Sciences
Korea Institute for Advanced Studies
Seoul, 130-722, Korea
Fax: (+82)2-958-3820
E-mail: danil@kias.re.kr

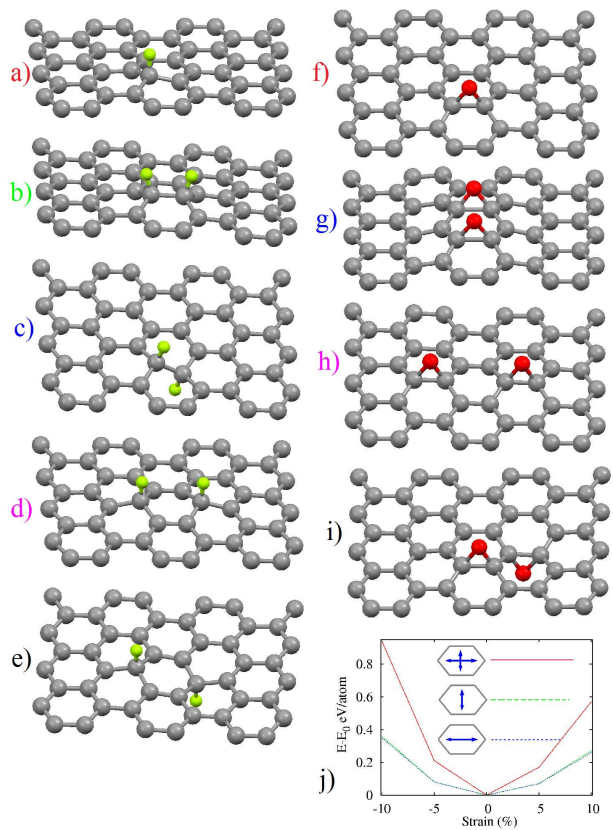

Figure 1. Optimized atomic structure for the graphene supercell (gray spheres) with chemisorbed hydrogen (a-e) and oxygen (f-i) adatoms (small light or dark spheres respectively) in different positions. On panel (j) is shown changes of the total energy of pristine graphene as function of uniaxial strain along zigzag (doted line) and armchair direction (dashed line) and isotropic strain (solid line). The insets denote the sketches of studied distortions of carbon hexagons in graphene.

To check the role of the chemical composition of the adsorbed species, we performed calculations for the fluorine adatoms and hydroxyl groups placed on graphene in same positions as hydrogen (Figure1a-e). Results of the calculations for the case of graphene with positive uniaxial strain along the zigzag direction confirm the aforementioned scenario for the chemisorption on the expanded graphene, although the $E_{chem}$ here is always negative (Figure 3). We note that the important structural distortions by chemisorbed species[13,^14,^53] play insignificant roles owing to the increase of lattice parameters (inset of Figure 3b) flattening of the distortions. A previous study[54] using molecular dynamic simulation showed the similar behaviors.[54] The main difference between enhancement of chemical activity of expanded and corrugated[16] graphene originates from atomic structures under external mechanical perturbations. In the case of corrugated graphene, the enhanced curvature provides a significant increase in binding energy that is limited only by the stability of carbon-carbon bonds in

graphene.[55,^56] This is quite different from the case of expanded graphene where the strain of 5% leads dramatic diminishment of out-of-plane carbon distortion near impurity.[13] Expansion of graphene by strain beyond 10% will lead further flattening of the distortion near impurity and eventually limits a further gain in binding energy.

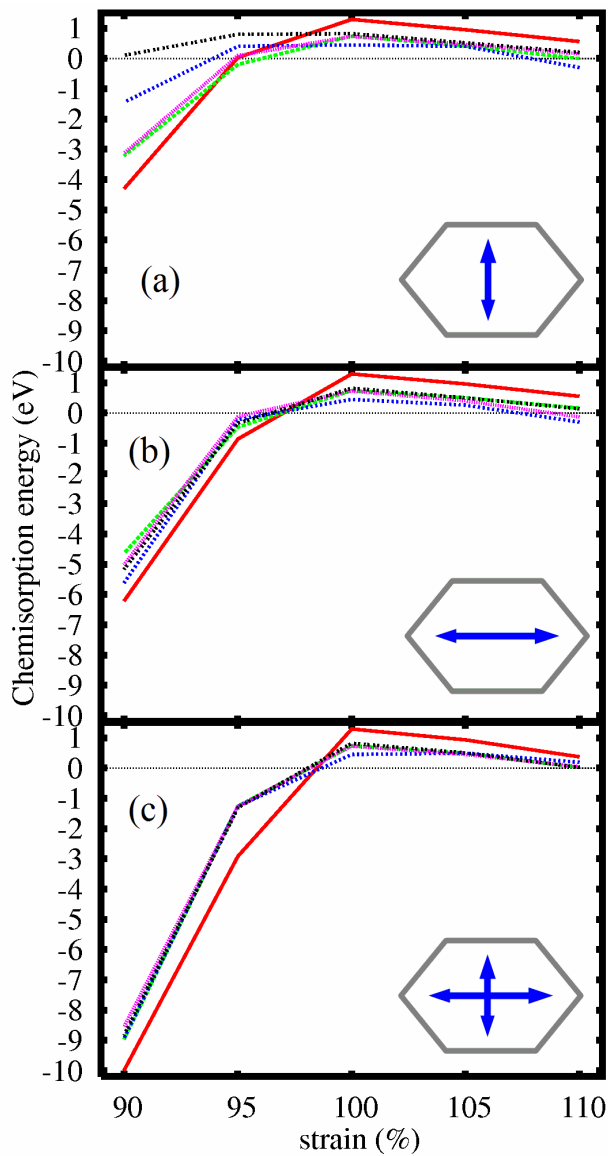

Figure 2. Chemisorption energies per hydrogen atom for the single hydrogen adatom (solid line) on graphene (see Fig. 1a) and its pair on various different positions (dashed for Fig. 1b, dotted for Fig. 1c, fine dotted for Fig. 1d and dashed black for Fig. 1e lines) under uniaxial strain along (a) armchair direction, (b) zigzag direction and (c) isotropic strain.

For compressed graphene, however, the situation changes drastically and local distortions around chemisorbed species start to play a crucial role in determining chemical activities. In the case of initially buckled by ripples[16] or negative strain[49] impurities chemisorbed by graphene lead to stabilization of initial corrugations. Our calculations demonstrate that chemically adsorbed species leads distortions of the initially flat graphene with negative strains turning carbon substrate into the energetically favoured state (see inset on Figures 3a and 4a-d). Like graphene with positive strain, increase in the total energy of compressed graphene as increasing the magnitude of the negative strains (Figure 1j) significantly decreases the chemisorption energy (Figures 3 and 4). For the case of the hydrogenation, it turns the process from the endothermic for pristine graphene to the exothermic for compressed graphene at a negative strain of around <M->5^%.

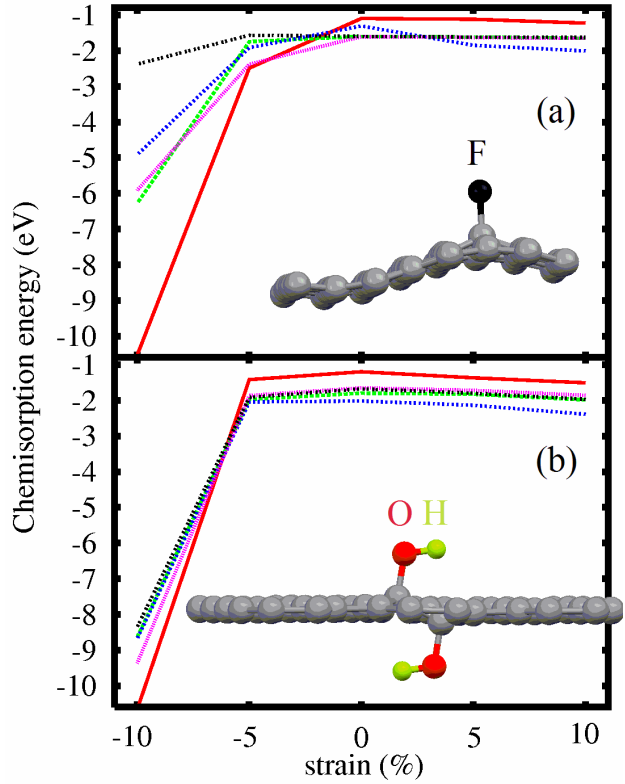

Figure 3 Chemisorption energies per (a) fluorine atom or (b) hydroxyl group for the single species (solid line) on graphene (see Fig. 1a) and its pair on various different positions (dashed for Fig. 1b, dotted for Fig. 1c, fine dashed for Fig. 1d and dashed black for Fig. 1e lines) under the uniaxial strain perpendicular to armchair direction. The insets show the optimized atomic structure of singe fluorine adatom on graphene compressed by 5% (a) and the pair of hydroxyl groups in *ortho* position from both side of graphene expanded by 10% (b).

**Magnetic properties of functionalized strained graphene**

The most surprising and important result for the chemisorptions on compressed graphene is the change in the energetically favourable atomic configuration for the pair of the species on both sides of graphene; from the paired chemisorbed species (the *ortho* position on both sides) to the isolated single chemisorbed species (Figure1a, inset of Figure 3a and Figure 4a--d). Hence, the paired adsorptions are not preferred in the compressed graphene. The cause of this surprising effect is the different distortions of the carbon hexagons provided by the chemisorption of single (Figure 4a, b) and pair (Figure 4c, d) hydrogen adatoms respectively. Single chemisorbed hydrogen adatom on graphene makes the two sublattices of hexagonal lattice non-equivalent. This causes the uniform alteration of the small and large hexagons in honeycomb lattice, similar to that discussed in ref. 19, if graphene is compressed (see top view on Figure 4b). Chemisorption of the second hydrogen atom on the carbon atom makes both sublattices equivalent again and destroys the uniform alteration. Therefore, the formation of the uniform distorted structure in the compressed graphene provides a significant energy gain for the case of single impurities chemisorbed on graphene.

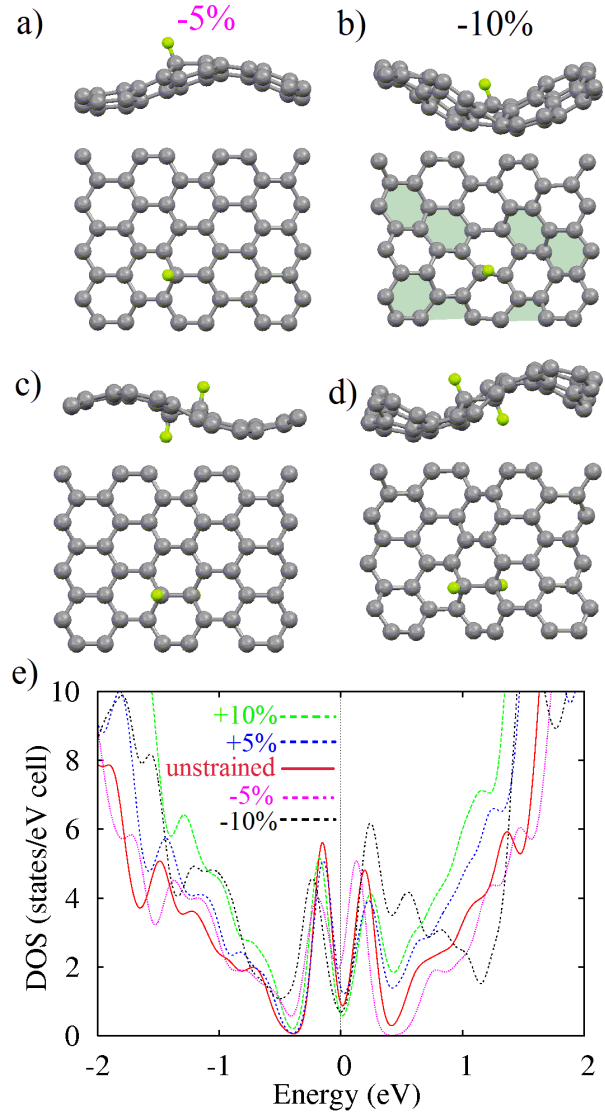

Figure 4. Optimized atomic structure for the graphene supercell (gray spheres) with a chemisorbed single (a, b) or pair (c, d) hydrogen adatoms (small light spheres) for the case of the negative strain of 5% (a) and of 10% (b). On panel (e) total densities of states near the Fermi energy (here we set it zero) for the strained graphene with a chemisorbed single hydrogen adatom under different types and magnitude of strains.

For the case of fluorine adatoms and hydroxyl groups on graphene the situation is similar to the hydrogen adsorptions on compressed graphene (Figure 5a). Bigger lattice distortions formed by the fluorine adatoms[13,45] make the chemisorption of a single fluorine adatom more favourable than other non-magnetic pairs. The magnetic ground states will not be destroyed by the pair formations of fluorine adatoms so that the compressed graphene is more feasible for the building of the magnetic graphene. The electronic structure near Fermi level for the case of single adatom chemisorption changed negligibly for all the kinds of strains. This stability near the Fermi level corresponds to the stability of magnetism. The magnetic moments change within 3% due to changes of the type and magnitude of applied strain and remain near 0.95 $\mu_B$ for a single adatom. Comparison of the energies between magnetic and nonmagnetic configurations (see insets of Figure 5) clearly shows the stability of the magnetism based on chemisorption of all studied species.

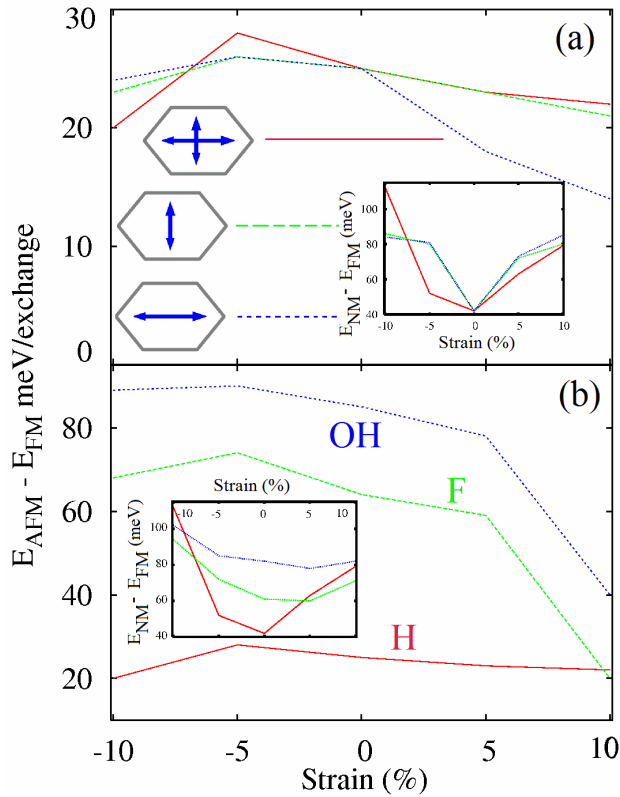

Figure 5. Energy difference between antiferromagnetic and ferromagnetic configurations of hydrogenated graphene (a) as function of the strain value for distortions along zigzag direction (dashed line), armchair direction (dotted blue lie) and isotropic strain (solid line); and energy difference between antiferromagnetic and ferromagnetic configurations for hydrogenated (solid line), fluorinated (dashed line) graphene and graphene with single hydroxyl group (dotted line) as functions of uniaxial strain (b). On insets are reported energy differences between paramagnetic and magnetic structures.

Fluorine-based magnetism is a rather controversial topic. Recent DFT calculations have shown the absence of a magnetic moment on single fluorine atoms[57] while the experimental results demonstrate the presence of paramagnetism in fluorinated graphene.[58] For detailed investigation of this issue we performed calculations for the case of a single fluorine adatom chemisorbed over graphene supercells with different sizes and shapes for different values of energy cut-off (from 100 to 600 Ry), different core radii of pseudopotentials (from 1.2 to 2.2 Å) and different $k$-point sampling (from 1×1×1 to 12×12×4). For smaller (from 8 to 18 carbon atoms) and bigger (above 72 carbon atoms) supercells self-consistent calculations always provide magnetic solution. For a medium-sized supercell, however, the self-consistent pseudopotential calculation leads to dependence of the magnetic solution and the value of the magnetic moment per supercell from the $k$-point sampling, while the full-potential method implemented in the ELK code[59] shows the opposite. We think that a possible computational error in determining an optimal bond length between carbon and fluorine (1.44 Å instead 1.40~1.43 Å in organic compounds)[60] influences its ground-state properties because, for the hydrogen-doped case, such a dependence of magnetic solution on $k$-points sampling does not occur.

To calculate exchange interaction parameters for the case of a single hydrogen impurity, we multiply the size of the supercell with the optimized atomic structure along the zigzag direction and calculated the total energy for the parallel and antiparallel orientations of the spins of two magnetic impurities.[27] The results of the calculations are shown in Figure 5a. Expansion of graphene provides a decrease of the exchange energy because the distance between impurities increases while a 5% compression of graphene increases the ferromagnetic exchange energy. Further compression over 5% changes the electronic structure (decreases the number of states at the Fermi level) and weakens the ferromagnetic interactions. A maximal value of exchange energy obtained for the uniaxial negative strain of -5% is used for the further discussion of thermal stability of magnetism in these systems. For the case of a single fluorine and hydroxyl group, a larger out-of-plane distortion of the graphene sheet provides enhancement of ferromagnetic interactions (Figure 5b).

## Lattice distortions and concentration effects

The aforementioned lattice distortions of the compressed graphene with chemisorption of various species require a detailed study of the lattice properties of compressed graphene. Large values for the chemisorption energies are caused by the significant energy gain of distorting the compressed flat graphene (Figure 1j). Hence, naively one can expect that such energy gain will indefinitely increase if the size of supercell is increased. However, the shape of lattice distortions strongly depends on the concentration of impurities and supercell size. To examine the role of the size of supercell we have performed the calculation on graphene uniaxially compressed by 10% with different sizes of the supercells and with an initial out-of-plane shift of several atoms at the center of supercell. In contrast to the initially flat graphene which remains flat after compression, this initial kick provides significant lattice distortions with different shapes (Figure 6a-d) depending on the sizes of supercell. To check whether the obtained structure is unique or not, we performed optimization of atomic structure for the case of chemisorption of a single hydrogen adatom on these supercells of compressed graphene. The chemisorption of the hydrogen provides the different distortion of graphene sheet (for example, see Figure 6b and Figure 4b). For the modelling of hydrogen desorption we started from these optimized atomic structures by increasing the carbon--hydrogen distance from 1.10 to 3 Å and performed optimization of these new structures. Obtained structure of distorted compressed graphene after hydrogen desorption is exactly the same as calculated for the case of pure compressed graphene with an initial out-of-plane shift of a few atoms in the center of supercell.

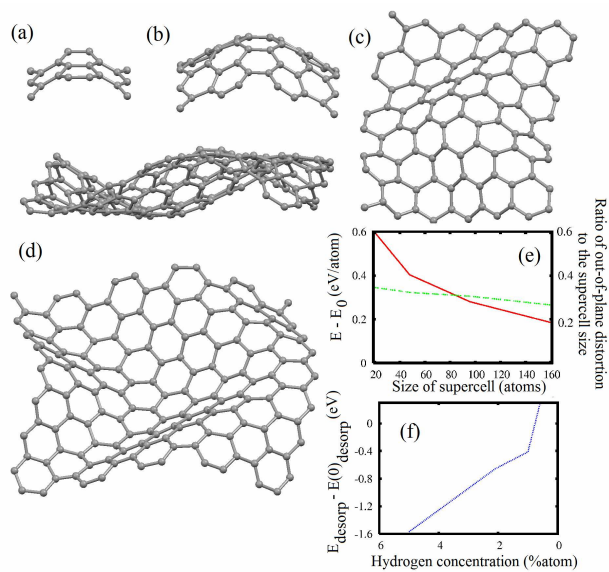

Figure 6 Optimized atomic structure of uniaxial compressed by 10% graphene with supercell contained 20 (a), 48 (b), 96 (c) and 160 (d) carbon atoms for the case of optimization of strained graphene started with initial out-of-plane shift of several carbon atoms in the centre of supercell. Changes of the total energy of pristine graphene as function of size compressed supercell (solid line) and ratio of the distortion of out of plane distortion to the width of supercell (dashed line) are shown on panel (e). Changes of the difference desorption energy of single hydrogen adatom on strained graphene and desorption energy for the single hydrogen on unstrained flat graphene as function of hydrogen concentration (f).

The total energy per carbon atom for the studied graphene supercells demonstrates that the energy gain for the distorted graphene covalent functionalization decreases as the system size increases (Figure 6e). These changes of the total energy are caused by the relative flattening of the graphene sheet that could be characterised by a reduction of the ratio of out-of-plane distortions to the size of the supercell (Figure 6e). Decay of graphene sheet out-of-plane distortions should provide decrease of the graphene chemical activity.[16] Comparison of hydrogen desorption energy of compressed graphene with unstrained graphene (Figure 6f) demonstrate that a single hydrogen atom chemisorbed on compressed graphene supercell with 160^^carbon atoms (0.6^% hydrogen concentration) is less stable than unstrained graphene. Thus we can conclude that giant chemisorption energy gains shown in Figures 2 and 3 are possible only for a concentration of impurities above 1%. In realistic conditions, thermal fluctuations (see ref. 54 and references therein) will play a role of initial out-of-plane kick and decrease chemisorption energies for the 10% compressed graphene by the values about 5 eV per hydrogen adatom.

## Oxidation of strained graphene

Unlike hydrogen, fluorine and hydroxyl groups oxygen adatoms on graphene form bonds with two nearest atoms with formation of epoxy groups (Figure 1f-i). This type of chemisorption provides several new features of strained graphene oxidations. Similar to the aforementioned chemical species the strain decreases the chemisorption energy (Figure 7). However, in sharp distinction to the hydrogen case, positive strain turns the oxidation process from endothermic to exothermic. In other words, graphene with positive strain could be easily oxidized in the air. Hydrogenation of strained graphene is not dependent on the direction of strains because the C-H bond is always perpendicular to the strain direction.

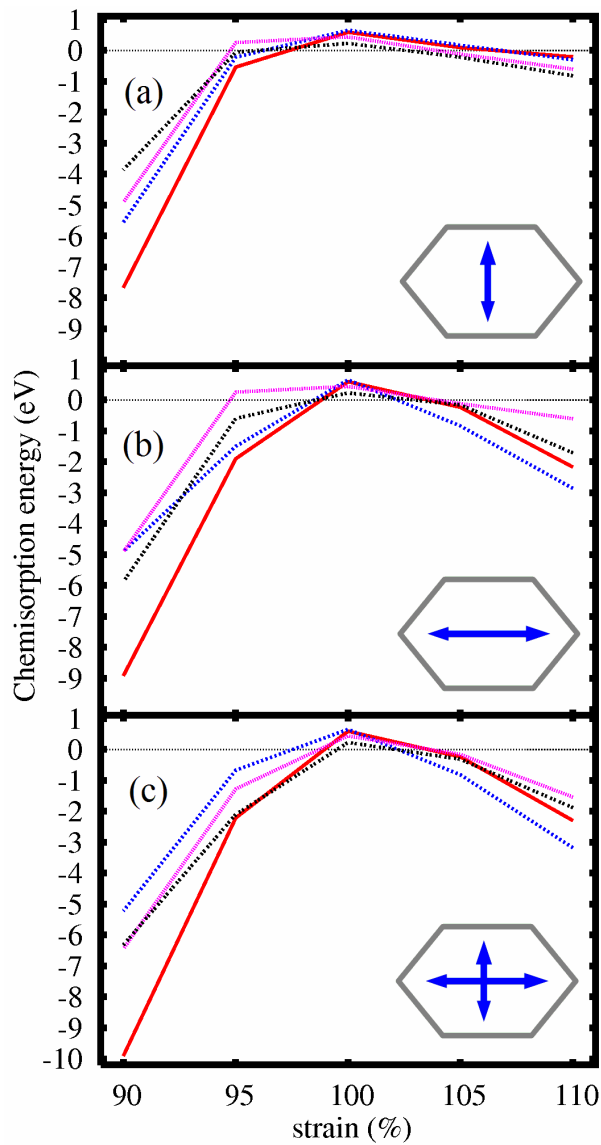

Figure 7. Chemisorption energies per oxygen atom for the single oxygen adatom (solid line) on graphene (see Fig. 1f) and its pair on various different positions (dotted for Fig. 1g, fine dotted for Fig. 1h and dashed black for Fig. 1i lines) for the uniaxial strain perpendicular to (a) armchair and (b) zigzag direction and (c) isotropic strain.

Oxidation, however, strongly depends on the strain direction because the lattice distortion by the formation of epoxy groups leads bigger expansion of the carbon hexagons in the zigzag direction.[14] For the case of unstrained graphene the oxygen pair shown in Figure 1i is energetically favourable and formation of the pair in Figure 1g requires the highest chemisorption energy. But expansion along the zigzag direction or compression perpendicular to the zigzag direction compensates anisotropic lattice distortions and makes formation of the pair shown on Figure 1g the most energetically favourable. Therefore the strain could be used for the selection of the initial configurations of the pair adsorptions so that different types of the oxygen lines along different directions on graphene can be formed. This could be used for selection of unzipping[61] or bending[62] directions of graphene by the oxidation. We expect that the natural lattice distortions of the graphene layers in carbon nanotubes also lead to rush a direction dependent oxidation resulting in oxidative unzipping of the nanotubes.[63,64]

## Conclusion

Based on first-principles calculations, we have studied hydrogenation, fluorination, and oxidation processes in strained graphene. By checking the variation of formation energies of a single adatom or its pairs upon applications of strains, the most probable chemical pathway to functionalization is deduced. When initially flat graphene is compressed, the single hydrogen and fluorine adatom configurations are the most favorable of any pair formation, enhancing the magnetic ground state against the destruction by pairing adatoms. Moreover, the oxidation process is unique in that the pair is preferred over a single adatom under strain and that the pair formation depends on the strain direction.

## Computational method

The modelling was carried out by first-principles pseudopotential calculation methods as implemented in SIESTA,[65] as was done in our previous works.[13,14,16,17,26,27,50,52] All calculations are performed using the generalized gradient approximation (GGA-PBE)[66] for exchange-correlation potential, which is more suitable for the description of graphene-adatom chemical bonds than local density approximation.[13] All atomic positions were fully optimized until the inter-atomic force is less than 0.04 eV/Å. Interlayer distance between graphene sheets have been chosen 20Å that provide perfect separation of the layers and makes unnecessary optimization of lattice parameter along Z axe. A rectangle-like supercell containing 48 carbon atoms (Fig. 1) in the graphene sheet has been used. We confirm that the size and shape of supercell are enough to guarantee the absence of any overlap between wave functions of the chemisorbed groups[27] and are feasible for the modelling of graphene sheet expansion and compression in zigzag and armchair directions (see insets on Fig. 2). When we performed the calculations for the strain along the one direction we optimize lattice parameters in the perpendicular direction. The lattice distortions after chemisorption of hydrogen and hydroxyl groups on pristine graphene are the same as obtained in our previous works for other sizes and shapes of supercells.[13,14] All calculations were carried out for an energy mesh cut off of 360 Ry and $k$-points of 8×8×1 Monkhorst-Pack mesh.[67]

The chemisorption energy has been calculated by a standard formula $E_{chem} = (E_{GX} - (E_G + E_X))/n$, where $E_{GX}$ is the total energy of the strained graphene with chemisorbed X species, $n$ a number of X, $E_G$ the total energy of strained pure graphene, and $E_X$ the total energy per species X (X = hydrogen, fluorine and oxygen) as the half of total energy of $H_2$, $F_2$ and $O_2$ molecule respectively. The total energy of the oxygen molecule calculated in ground (triplet) state. The energy of the hydroxyl group is defined as $E_{OH} = (2E_{H2O} - E_{H2})/2$ where $E_{H2O}$ and $E_{H2}$ are total energies of $H_2O$ and $H_2$ respectively. Desorption energy of single hydrogen adatom we define as $E_{desorp} = E_{GH} - E_{G+H}$, where $E_{GH}$ - energy of optimized atomic structure of graphene supercell with chemisorbed hydrogen atom, and $E_{G+H}$ - energy of the optimized atomic structure of the same supercell after ablation of hydrogen adatom to the distance about 3Å from nearest carbon atom.

## Acknowledgements

We acknowledge computational support from the CAC of KIAS. Y.-W. S. was supported by the NRF grant funded by MEST (Quantum Metamaterials Research Center, R11-2008-053-01002-0 and Nano R&D program 2008-03670).

**Keywords:** graphene, strain, first-principles calculations, functionalization, sp-magnetism

[1]     A. K. Geim, K. S. Novoselov, *Nat. Mater.* **2007**, *6*, 183-91.

[2]     M. I. Katsnelson, *Mater. Today* **2007**, *10*, 20-7.

[3]     A. H. Castro Neto, F. Guinea, N. M. R. Peres, K. S. Novoselov, A. K. Geim, *Rev. Mod. Phys.* **2009**, *81*, 109-62.

[4]     A. K. Geim, *Science* **2009**, *324*, 1530-4.

[5]     N. Tombros, C. Jozsa, M. Popinciuc, H. T. Jonkman, B. J. van Wees, *Nature* **2007**, *448*, 571-4.

[6]     O. V. Yazyev, M. I. Katsnelson, *Phys. Rev. Lett.* **2008**, *100*, 047209.

[7]     K. P. Loh, Q. L. Bao, G. Eda, M. Chhowalla, *Nature Chem.* **2010**, *2*, 1015-24.

[8]     P. Blake, P. D. Brimcombe, R. R. Nair, T. J. Both, D. Jiang, F. Scheidin, L. A. Ponomarenko, S. V. Morozov, H. F. Gleeson, E. W. Hill, A. K. Geim, K. S. Novoselov, *Nano Lett.* **2008**, *8*, 1704-8.

[9]     S. R. C. Vivekchand, C. S. Rout, K. S. Shubrahmanyam, A. Govindraj, C. N. R. Rao, *J. Chem. Sci.* **2008**, *120*, 9-13.

[10]    H. Wang, Q. Hao, X. Yang, L. Lu, X. Wang, *Electrochm. Commun.* **2009**, *11*, 1158-61.

[11]    D. Yu, L. Dai, *J. Phys. Chem. Lett.* **2010**, *1*, 467-70.

[12]    S. Chen, J. Zhu, X. Wu, Q. Han, X. Wang, *ACS Nano* **2010**, *4*, 2822-30.

[13]    D. W. Boukhvalov, M. I. Katsnelson, *J. Phys.: Condens. Matter* **2009**, *19*, 344205.

[14]    D. W. Boukhvalov, M. I. Katsnelson, *J. Am. Chem. Soc.* **2008**, *130*, 10697-702.

[15]    I. Jing, D. A. Dikin, R. D. Piner, R. S. Ruoff, *Nano Lett.* **2008**, *8*, 4283-7.

[16]    D. W. Boukhvalov, M. I. Katsnelson, *J. Phys. Chem. C* **2009**, *113*, 14176-8.

[17]    D. C. Elias, R. R. Nair, T. G. M. Mohiuddin, S. V. Morozov, P. Blake, M. P.; Halsall, A. C. Ferrari, D. W. Boukhvalov, M. I. Katsnelson, A. K. Geim, K. S. Novoselov, *Science*, **2009**, *323*, 610.

[18]    Z. Luo, T. Yu, K.-J. Kim, Z. Ni, Y. You, S. Lim, Z. Shen, S. Wang, J. Lin, *ACS Nano* **2009**, *3*, 1781.

[19]    M. Jaiswal, C. H. Y. H. Lim, Q. L. Bao, C. T. Toh, K. P. Loh, B. Ozylmaz, *ACS Nano* **2011**, *5*, 888-896.

[20]    S.-H. Cheng, K. Zhouu, F. Okino, H. R. Gutierrez, A. Gupta, N. Shen, P. C. Eklund, J. O. Sofo, J. Zhu, *Rhys. Rev. B* **2010**, *81*, 205435.

[21]    R. R. Nair, W. Ren, R. Jalil, I. Riaz, V. G. Kravets, L. Britnell, P. Blake, F. Schedin, A. S. Mayorov, S. Yuan, M. I. Katsnelson, H.-M. Cheng, W. Strupinski, L. G. BulushevaA. V. Okotrub, I. V.; Grigorieva, A. N. Grigorenko, K. S. Novoselov A. K. Geim *Small* **2010**, *6*, 2877-84.

[22]    M. Sepioni, S. Rablen, R. R. Nair, J. Narayanan, F. Tuna, R. Winpenny, A. K. Geim, I. V. Grigorieva, *Phys. Rev. Lett.* **2010**, *105*, 207205.

[23]    O. V. Yazyev, *Rep. Prog. Phys.* **2010**, *73*, 056501.

[24]    Y. Wang, Y. Hang, Y. Song, X. Zhang, Y. Ma, J. Liang, Y. Chen, *Nano Lett.* **2009**, *9*, 220-4.

[25]    J. M. Hong, S. Niyogi, E. Bekyarova, M. E.; Itkis, P. Ramesh, N. Amos, D. Litvinov, C. Berger, W. A. de Heer, S. Kirzoev, R. C. Haddon, *Small* **2011**, *7*, 1175-80.

[26]    D. W. Boukhvalov, M. I. Katsnelson, *ACS Nano* **2011**, *5*, 2440-6.

[27]    D. W. Boukhvalov, Katsnelson, M. I. *Nano Lett.* **2008**, *8*, 4378.

[28]    S. Stankovich, D. A. Dikin, G. H. B. Dommett, K. M. Kolhaas, E. J. Zimney, E. A. Stach, R. D. Piner, S. T. Nguyen, R. S. Ruoff, *Nature* **2006**, *442*, 282-6.

[29]    A. V. Talyzin, S. M. Luzan, T. Szabo, D. Chernyshev, V. Dmitriev, *Carbon* **2011**, *49*, 1894-9.

[30]    S. Park, R. S.; Ruoff, *Nature Nanotech.* **2009**, *4*, 217;

[31]    M. J. Allen, V. C. Tung, R. B. Kaner, *Chem. Rev.* **2010**, *110*, 132;

[32]    D. R. Dreyer, S. Park, C. W. Bielawski, R. S. Ruoff, *Chem. Soc. Rev.* **2010**, *39*, 228.

[33]  E. Bekyarova. M. E.; Itkis. P. Ramesh, R. C. Haddon, *phys. stat. solidi – RRL* **2009**, *3*, 184-6.

[34]  J. C. Meyer, A. K. Geim, M. I. Katsnelson, K. S. Novoselov, T. J. Booth, S. Roth, *Nature* **2007**, *446*, 60-3.;

[35]  E. Stolyarova, K. T. Rim, S. Ryu, J. Maultzsch, P. Kim, L. E. Brus, T. F. Heinz, M. S. Hybertsen, G. W. Flynn, *Proc. Natl. Acad. Sci. U.S.A.* **2007**, *104*, 9209-12;

[36]  M. Ishigami, J. H. Chen, W. G. Cullen, M. S. Fuhrer, *Nano Lett.* **2007**, *7*, 1643-8.

[37]  A. Fasolino, J. H. Los, M. I. Katsnelson, *Nat. Mater.* **2007**, *6*, 858-61.

[38]  Y.-W. Son, M. L. Cohen, S. G. Louie, *Nature* **2006**, *444*, 347.

[39]  Y.-W. Son, M. L. Cohen, S. G. Louie, *Phys. Rev. Lett.* **2006**, *97*, 216803.

[40]  F. Guinea, M. I. Katsnelson, A. K. Geim, *Nature Phys.* **2009**, *6*, 30-3.

[41]  N. Levy, S. A. Burke, K. L. Meaker, M. Panlasigui, A. Zettl, F. Guinea, A. H. Castro Neto, M. F. Crommie, *Science* **2010**, 329, 544.

[42]  S.-M. Choi, S.-H. Jhi, Y.-W. Son, *Phys. Rev. B* **2010**, 81, -81407(R);

[43]  M. A. H. Vozmediano, M. I. Katsnelson, F. Guinea, *Phys. Rep.* **2010**, 496, 109.

[44]  V. M. Pereira, A. H. Castro Neto, *Phys. Rev. Lett.* **2009**, 103, 046801.

[45]  M. Yang, A. Nurbawono, C. Zhang, R. Wu, Y. Feng, Ariando Ariando *AIP Advances* **2011**, *1*, 032109.

[46]  M. Zhou, Y. Lu, C. Zhang, Y.-P. Feng, *Appl. Phys. Lett.* **2010**, *97*, 103109.

[47]  K. Xue, Z. Xu, *Appl. Phys. Lett.* **2010**, *96*, 063103;

[48]  Y. Ma, Y. Dai, C. Niu, L. Yu, B. S. Huang, *Nanoscale* **2011**, *3*, 2301-6.

[49]  Z. F. Wang, Y. Zhang, F. Liu, *Phys. Rev. B* **2011**, *83*, 041403.

[50]  D. W. Boukhvalov, X. Feng, K. Müllen, *J. Phys. Chem. C* **2011**, *115*, 16001-5.

[51]  J. O. Sofo, A. M. Suarez, G. Usaj, P. S. Cornaglia, A. D. Hernández-Nieves, C. A. Balseiro, *Phys. Rev. B* **2011**, *83*, 081411.

[52]  D. W. Boukhvalov, *Physica E* **2010**, *43*, 199.

[53]  D. W. Boukhvalov, *Nanotechnology* **2011**, *22*, 055708.

[54]  R. Roldán, A. Fasolino, K. Zakharchenko, M. I. Katsnelson, *Phys. Rev. B* **2011**, *83*, 174104.

[55]  C. Lee, X. Wei, J. W. Kysar, J. Hone, *Science* **2008**, *321*, 385-8;

[56]  C. A. Marinetti, H. G. Yevick, *Phys. Rev. Lett.* **2010**, *105*, 245502.

[57]  G. Usaj, P. S. Cornaglia, A. D. Hernández-Nieves, C. A. Balseiro, J. O. Sofo (unpublished).

[58]  R. R. Nair, M. Sepioni, I.-L. Tsai, O. Lehtinen, J. Keikonen, A. V. Krasheninnikov, T. Thomson, A. K. Geim, I. V. Grigorieva arXiv:1111.3775.

[59]  K. Dewhurst, S. Sharma, L. Nordström, F. Cricchio, F.Bultmark, Hardy Gross, C. Ambrosch-Draxl, C. Persson, C. Brouder, R. Armiento, A. Chizmeshya, P. Anderson, I. Nekrasov, F. Wagner, F. Kalarasse, J. Spitaler, S. Pittalis, N.Lathiotakis, T. Burnus, S. Sagmeister, C. Meisenbichler, S. Lebègue, Y. Zhang, F. Körmann, A. Baranov, A. Kozhevnikov, S. Suehara, F. Essenberger, A. Sanna, T. McQueen, T. Baldsiefen, M. Blaber, A. Filanovich, T. Björkman, M. Stankovski, J. Goraus, M. Meinert, D. Rohr, V. Nazarov, Elk-1.4.5 code. http://elk.sourceforge.net/

[60]  F. H. Allen, O. Kennard, D. G. Watson, L. Brammer, A. G. Orpen, *J. Chem. Soc. Perkin Trans.* **1987**, S1-S19.

[61]  S. Fujii, T. Enoki, *J. Am. Chem. Soc.* **2010**, *132*, 10034.

[62]  J.-L. Li, K. N. KudinM. J. McAllister, R. K. Prud'homme, I. Aksay, R. Car, *Phys. Rev. Lett.* **2006**, *96*, 176101.

[63]  D. V. Kosynkin, A. L. Higginbotham, A. Sinitskii, J. R. Lomeda, A. Dimiev, B. K. Price, J. M. Tour, *Nature* **2009**, *458*, 872;

[64]  L. Jiao, L. Zhang, X. Wang, G. Diankov, H. Dai, *Nature* **2009**, *458*, 877.

[65]  J. M. Soler, E. Artacho, J. D. Gale, A. Garsia, J. Junquera, P. Orejon, D. Sanchez-Portal, *J. Phys.: Condens. Matter* **2002**, *14*, 2745-79.

[66]  J. P. Perdew, K. Burke, M. Ernzerhof, *Phys. Rev. Lett.* **1996**, *77*, 3865-8.

[67]  H. J. Monkhorst, J. D. Park, *Phys. Rev. B* **1976**, *13*, 5188-92.